\documentstyle[12pt]{article}
\begin{document}
\baselineskip=1.3\baselineskip
\baselineskip=1.2\baselineskip
\renewcommand{\thefootnote}{\fnsymbol{footnote}}
\setcounter{equation}{0}
\newcommand{\beq}{\begin{equation}}
\newcommand{\eeq}{\end{equation}}
\newcommand{\beqa}{\begin{eqnarray}}
\newcommand{\eeqa}{\end{eqnarray}}
\pagestyle{plain}
\begin{titlepage}

\hfill{ ANL-HEP-PR-95-46; IP/BBSR/95-62}
\vspace{0.6cm}
\begin{center}
{\large{\bf  {The Electric Charge of a Dirac Monopole \\}
{ at Nonzero Temperature}}}\par
\end{center}
\vskip 1.0cm
\begin{center}
{Claudio Corian\`{o}$^{a,}$\footnote{email :
coriano@hep.anl.gov.~ Work supported by
the U.S.~Department of Energy, Division of High Energy Physics, Contract
W-31-109-ENG-38.}\ and \
Rajesh R. Parwani$^{a,b,}$\footnote{email : parwani@iopb.ernet.in}}
\end{center}
\vskip 0.6cm
\centerline{$^a$High Energy Physics Division}
\centerline{Argonne National Laboratory}
\centerline{9700 South Cass, Il 60439, USA.}
\vspace{0.5cm}
\centerline{$^b$Institute of Physics, Bhubaneswar 751005, India.}

\vskip 1.0cm
\centerline{June 1995}
\vskip 1.0 cm
\centerline{\bf Abstract}
We study the effect of nonzero temperature on the induced electric charge
around
a Dirac monopole. While at zero temperature the charge is known to be
proportional to a CP violating $\theta$ parameter, we find that at high
temperature the charge is proportional to sin  $\theta$. Other features
of the charge at nonzero temperature are discussed. We also compute
the induced charge at nonzero temperature around an Aharonov-Bohm flux
string in $2+1$ dimensions and compare the result with an index theorem,
and also with the electron-monopole problem in $3+1$ dimensions.

\end{titlepage}
Dirac \cite{Dirac} showed that the quantum mechanics of an electron in the
presence of a fixed magnetic monopole was consistent only if the electron
charge $e$ and the monopole magnetic charge $g$ satisfied the quantization
condition
\beq
q\equiv e g ={1\over 2} \times \mbox{integer}.
\label{one}
\eeq

While Dirac described his monopole in terms of a singular vector
potential, a non-singular description was later obtained \cite{WY}
which gave a simpler derivation of (\ref{one}).
The non-singular formulation has the advantage of enabling a clearer
discussion of the dynamics of a fermion interacting with a monopole
\cite{KYG,G,Y}. A careful investigation of the fermion-monopole
problem \cite{G} revealed that the Hamiltonian of the system was not
self-adjoint, leading thereby to non-real eigenvalues.
By imposing appropriate boundary conditions on the wavefunctions,
the Hamiltonian could  be made
self-adjoint, but at the price of introducing an undetermined
angular $\theta$ parameter.
The classical $CP$ symmetry of the problem is maintained only for the
particular choices $\theta=0$ or $ \pi$ \cite{G}.

For a general $\theta$, it was found \cite{Y} that the monopole acquired
an electric charge due to vacuum polarization. The electric charge
was computed to be
\beq
Q= -{e \theta\over 2 \pi} 2 q
\label{two}
\eeq
in agreement with a general statement of Witten \cite{EW}.

Our purpose here is to investigate how the relation (\ref{two}) is modified
at non-zero temperature. We consider a fixed Dirac monopole in an
electron-positron plasma at temperature $\beta=1/T$. That is, we ignore
first the effects of  photons which will be estimated later.

In terms of the eigenstates of the electron-monopole Hamiltonian,
the thermal ensemble average of the induced charge is given by
\beq
Q= {-e\over 2}\int d^3 {\bf r} \sum_{E,j,m} \ \tanh \,{\beta |E|
\over 2} \ \mbox{sign}(E) \
\psi^{\dag}_{E,j,m}\psi_{E,j,m} \; .
\label{three}
\eeq
In (\ref{three}) $E$ is the energy, $j$ labels the total angular momentum
and $m$ labels the $z-$component of angular momentum.
The values of $j$ increase in steps of one \cite{WY,KYG}
starting from the lowest value
$j_0=|q|-1/2$. For any $j\ge j_0$ there is a norm-preserving map \cite{Y}
between
the negative and the positive energy eigenstates so that only the
lowest partial wave $j_0$ contributes to (\ref{three}).
Using the expressions for the wave functions given in \cite{Y} we obtain
\beq
Q=Q_b + Q_c,
\eeq
where
\beqa
&& Q_b= -{e M\over 2}(2 q)(-2 \cos \theta ) \ \Theta(-\cos \theta) \
\mbox{sign}(\sin \theta) \ \tanh{\beta M|\sin\theta|\over 2} \
 \int_{0}^{\infty} dr \ e^{-2 r M|\cos\theta|} \nonumber \\
&& \label{five}
\eeqa
is the contribution from a bound state which exists if
$\cos\theta<1$, and
\beq
Q_c = -{e\over 2 \pi}(2 q)({-M\over 2})\int_{0}^{\infty} dr
\int_{-\infty}^{\infty} {dk\over \omega_k} \ \tanh {\beta\omega_k\over 2}
\ {2 i k \sin\theta \over M \cos\theta - i k} \ e^{2 i k r}
\label{six}
\eeq
is the contribution from the continuum states.
In the above, $M$ is the electron mass and $\omega_k\equiv \sqrt{k^2 + M^2}$.
The $k$ integral in (\ref{six}) may be evaluated by forming a contour
integral in the upper half of the complex $k-$plane. For $\cos\theta<1$
there will be a contribution to $Q_c$
from a pole at $k=-i M \cos\theta$ which cancels
exactly $Q_b$, just as at zero temperature.  Note that at non-zero
temperature there is no branch cut in the complex $k-$plane. There
is however
an infinite string of temperature dependent poles along the imaginary axis
due to the $\tanh\,(\beta\omega_k/2)$ factor. Evaluating the
contribution of these poles we obtain

\beq
Q=-{e x\over \pi}(2 q) \ \sin\theta \ \sum_{n=0}^{\infty}{1\over
(2 n +1)^2 + x^2 +x \cos\theta \ \sqrt{(2 n +1)^2 + x^2}} \; ,
\label{seven}
\eeq
where $x \equiv M/(\pi T)$. A number of features are apparent from
(\ref{seven}): (a) As expected, the charge decreases as the temperature
increases ($x\to 0$), (b) the expression is odd under
$\theta\to -\theta$ just as at zero temperature (so from now on we
will discuss only the range $0<\theta<\pi$),
(c) at a fixed temperature the charge decreases as $\theta$ increases from
$0$ to $\pi/2$, $(d)$ For $\theta$ between $\pi/2$ and $\pi$ the expression
(\ref{seven}) has no obvious universal behaviour other than vanishing
at $\theta=\pi$. Thus at nonzero temperature the charge vanishes at the
$CP$ even values of $\theta=0$ and $\pi$.

Consider the high temperature
$(x\to 0$) limit of (\ref{seven}). One easily obtains
\beqa
 Q &\to& -{e M\over T}(2 q) \ {\sin\theta\over \pi^2} \
\sum_{n=0}^{\infty}{1\over (2 n +1)^2} \ + \ O\left({M\over T}\right)^2
\nonumber
\\
&& = -{e M\over T}(2 q) \ {\sin\theta\over 8} \ + \ O\left({M\over T}\right)^2.
\label{eigth}
\eeqa
The most striking feature of  the high temperature limit is the
proportionality of the charge to $\sin\theta$ rather than $\theta$ which is
the case at zero temperature. Therefore at high temperature the charge
is a maximum at $\theta=\pi/2$.
The behaviour of the charge as a function of the $CP$ violation parameter
$\theta$ for different temperatures is shown in the figure. It is interesting
to note also from (\ref{six}) that the charge density at large distances
goes as
\beq
{\rho(r)} = O \left( {e^{-r \sqrt{M^2 +\pi^2 T^2}} \over r^2}\right) \, .
\label{den}
\eeq

Let us now show that in the limit of zero temperature ($x\to \infty$)
the expression (\ref{seven}) reduces to (\ref{two}).
Split the sum in (\ref{seven}) as $\sum_{n=0}^{N}\,\,+ \sum_{n=N}^{\infty}$
for some $N\gg 1$. Then for $x\gg N$, the finite sum contributes to $Q$ an
amount which vanishes as $O(1/x)$, while the contribution of the sum
$\sum_{n=N}^{\infty}$ may be estimated by writing it as an integral.
Thus
\beq
\lim_{x\to \infty} Q\sim \lim_{x\to \infty} -{e x \over \pi}(2 q)\sin\theta
\int_{\Lambda}^{\infty}{dy\over (2 y)^2 + x^2 + x \cos\theta
\sqrt{(2 y)^2 + x^2}},
\eeq
where $\Lambda\sim N$.
Making the change of variables $2 y=x \sinh z$ we obtain
\beqa
 \lim_{x\to\infty} Q &=&-{e\over 2 \pi}(2 q)\sin\theta \ \lim_{x\to\infty}
\int^{\infty}_{\sinh^{-1}\,{2\Lambda\over x}} \
{dz \over \cosh z + \cos\theta} \nonumber \\
&=&-{e\over 2 \pi}(2 q) \ \sin\theta \ {\theta\over \sin\theta}\nonumber \\
&=& -{e\theta\over 2 \pi}(2 q) \, ,
\eeqa
which is Eq.~(2).

There is one value of $\theta$ for which the expression (\ref{seven}) may be
evaluated in closed form for any temperature. For $\theta=\pi/2$ we
obtain
\beqa
 Q(\theta = \pi/2)&=&-{e x\over \pi}(2 q)  \ \sum_{n=0}^{\infty}
{1\over (2 n +1)^2 + x^2}\nonumber \\
&=& -{e\over 4}(2 q) \tanh {M\over 2 T}
\label{ten}
\eeqa
which agrees at high temperature with (\ref{eigth}) and at zero temperature
with (\ref{two}).
It is known \cite{HG} that for $\theta=\pi/2$, the problem of induced charge
around a monopole (with $2 q=1$) in 3+1 dimensions is mathematically
equivalent to the problem of induced charge around an Aharonov-Bohm (AB)
flux string
with flux $F=1/2$ in 2+1 dimensions. However the AB problem is exactly
solvable for any flux not only at zero temperature, but also at nonzero
temperature, as we now illustrate.

Using the wavefunctions for the electron-AB flux string system (see,
for example \cite{PG}), we obtain
the following expression for the induced charge at nonzero temperature
\beq
Q_{AB}=-{e M \mbox{sign}(F)\over 4 \pi}\int d^2{\bf r}\int_{0}^{\infty}
{k dk\over \omega_k} \ \tanh{\beta\omega_k\over 2} \
\left(J^2_{-|F|}(kr)-J^2_{|F|}(kr)\right) \, ,
\label{eleven}
\eeq
where $F=-{e\over 2 \pi}\oint\vec{A}\cdot d\vec{l}$ is the flux ($|F| < 1$)
in the
string, $r$ is the radial distance
from the string, and $J_\mu(kr)$ is the Bessel function.
Note the following inverse-Mellin-transform  representation for the
hyperbolic tangent,
\beq
\tanh{\beta\omega_k\over 2}={1\over 2 \pi i}\int_{c-i\infty}^{c+ i\infty}
ds \ \omega^{-s}_k \ \left[2 \beta^{-s}(2^{1-s}-1)\Gamma(s)\zeta(s)\right].
\label{twelve}
\eeq
Define also the integral
\beq
A_{\nu}(t)\equiv\int_{0}^{\infty} \alpha^{\nu-1}e^{-t \alpha^2} d\alpha .
\label{thirteen}
\eeq
A scaling of (\ref{thirteen}) gives
\beq
\omega^{-s-1}_k={A_{s+1}(\omega^2_k)\over A_{s+1}(1)}.
\label{fourteen}
\eeq
Using (\ref{twelve}) and (\ref{fourteen}) in (\ref{eleven}), the
$\int dk$ and the $\int d^2{\bf r}$ integrals become the same as at zero
temperature \cite{PG}, giving thereby
\beqa
Q_{AB}&=&-eM { \mbox{sign}(F)\over 4 \pi}{2\over 2 \pi i}
\int_{c-i\infty}^{c+ i\infty}ds \ \beta^{-s}(2^{1-s}-1)\Gamma(s)\zeta(s)
\ {2 \pi |F|\over |M|^{s+1}}\nonumber \\
&=&-e{F M \over 2 |M|}\ \tanh {\beta |M|\over 2} \nonumber \, .
\label{fifteen}
\eeqa
That is,
\beq
Q_{AB}=-e{F\over 2} \ \mbox{sign}(M) \ \tanh{\beta |M|\over 2}.
\label{17}
\eeq
As stated, $Q_{AB}(F=1/2)$ agrees with Eq.~(\ref{ten}) for $2q=1$.
(In (\ref{fifteen}) the $\mbox{sign}(M)$
refers to the irreducible representations
of two-component spinors in $2+1$ dimensions.
So the mapping here is for $M>0$). The calculation above for
the AB case can be extended beyond $|F| <1$ by taking into account the
contribution from a discrete set of threshold states \cite{PG}.

In $2 +1$ dimensions, the induced charge around any static external
magnetic flux $F$ is  a topological invariant at zero temperature
\cite{NS},  $Q_{2+1}(T=0)=-{e F\over 2} \mbox{sign}(M).$
It has been argued \cite{N} to be an invariant also at nonzero temperature
by using the fact that the regulated  index $\eta(s)\equiv
\sum_{E} \mbox{sign}(E) |E|^{-s}$ for the corresponding Dirac Hamiltonian is
an invariant (depending only on the total flux rather than the details of the
field). Indeed the induced charge due to
a uniform  external magnetic field in $QED_3$ is given by the right-hand-side
of
(\ref{fifteen}) and our explicit calculation of (\ref{fifteen}) for the
AB case verifies the argument of \cite{N}.

We mention now some corrections to the result (\ref{seven}).
The effect of gauge-field fluctuations (virtual and thermal photons)
on the magnitude of the
induced charge should be suppressed by powers of the fine structure
constant (this is certainly true for $2 q\gg 1$). From (\ref{den})
we see that the
induced charge is more localized around the monopole at nonzero temperature
than at zero temperature. Far from the monopole this electric charge
will be  screened by plasma collective effects; at high temperature the
Debye screening mass is $\sim e T$. On the other
hand the static magnetic field
of the monopole remains unscreened (for a plasma in its normal state).
If there is a non-zero chemical potential, then the relative contribution of
the positive and negative energy eigenstates to the charge is different,
so that in the electron-monopole problem
the states with $j\ge |q|+1/2$ will contribute.
(For $QED_3$ the induced charge around a magnetic flux tube is no longer
an invariant when the chemical potential is non-zero \cite{N}).
It should be noted also that the fractional charge
as computed from Eq.(\ref{three})
is a thermal  expectation value rather than a sharp eigenvalue
which would be the case  at zero
temperature \cite{KG}.\\

We conclude by summarizing the main features of the induced charge around
the monopole:

(i) The induced charge decreases with increasing temperature, going as
$\sim M/T$ at large temperatures.

(ii) The dependence of the charge on the $\theta$ parameter is modified
from that at zero temperature (see Figure). The charge vanishes at the
$CP$
even values of $\theta = 0$ and $\pi$. At high temperature the charge is
proportional to $\sin\theta$.

(iii) The induced charge becomes more localized with an increase in
temperature.
At high temperatures it is localized to within the thermal Compton wavelength
$1/T$.

(iv) At a non-zero temperature the charge vanishes in the limit of
massless fermions, just as is the case for induced charge in $QED_3$
(cf Eq.(\ref{17})
and \cite{N}).

An interesting open question is how the analysis of \cite{EW}, which holds
more generally
for extended monopoles \cite{HP}
in non-Abelian gauge theories with a CP violating vacuum angle,
 changes at non-zero temperature.

\centerline{Acknowledgements}
We thank Alfred Goldhaber for a reading of the manuscript and helpful
comments. R.P. also thanks the High Energy Theory Group at Argonne,
in particular  Alan White,
for hospitality during the course of this work.

\vspace{0.3cm}
\noindent{{\large{\bf Figure Caption}}\\
A plot of $S$ against $\theta$ for several values of $x=M/\pi T$. The charge is
given by $Q= -{e (2q) \over \pi} S$. The five curves correspond to the values
$x=0.05, \ 0.1, \ 0.5, \ 1$ and $10$,
the curves with increasing amplitude corresponding
to larger $x$ (lower temperatures). For the $x=10$ curve (which is almost a
straight line approaching the zero temperature relation (\ref{two})),
we have not indicated the sharp
drop which occurs near $\theta=\pi$.


\newpage
\begin{thebibliography}{99}
\bibitem{Dirac} P.A.M. Dirac, Proc. Roy. Soc. A133 (1931) 60.
\bibitem{WY} T.T. Wu and C.N. Yang, Phys. Rev. D12 (1975) 3845;
Nucl Phys B107 (1976) 365.
\bibitem{KYG} Y. Kazama, C.N. Yang and A.S. Goldhaber,
Phys. Rev. D15 (1977) 2287.
\bibitem{G} A.S. Goldhaber, Phys. Rev. D16 (1977) 1815.
\bibitem{Y} H. Yamagishi, Phys. Rev. D27 (1983) 2383.
\bibitem{EW} E. Witten, Phys. Lett. B86 (1979) 283.
\bibitem{HG} Hou Bo-yu et al., Comm. Theor. Phys. 7 (1987) 49;\\
Ph. De Sousa Gerbert, Phys. Rev. D40 (1989) 1346.
\bibitem{PG} R.R. Parwani and A.S. Goldhaber, Nucl. Phys. B359 (1991) 483.
\bibitem{NS} A.J. Niemi and G. Semenoff, Phys. Rep. 135 (1986) 99.
\bibitem{N} A.J. Niemi, Nucl. Phys. B251 (1985) 155.
\bibitem{KG} S. Kivelson and A.S. Goldhaber, Phys. Lett. B255 (1991) 445.
\bibitem{HP} G.'t Hooft, Nucl. Phys. B79 (1974) 276;\\
A. M. Polyakov, JETP Lett. 20 (1974) 194.
\end{thebibliography}
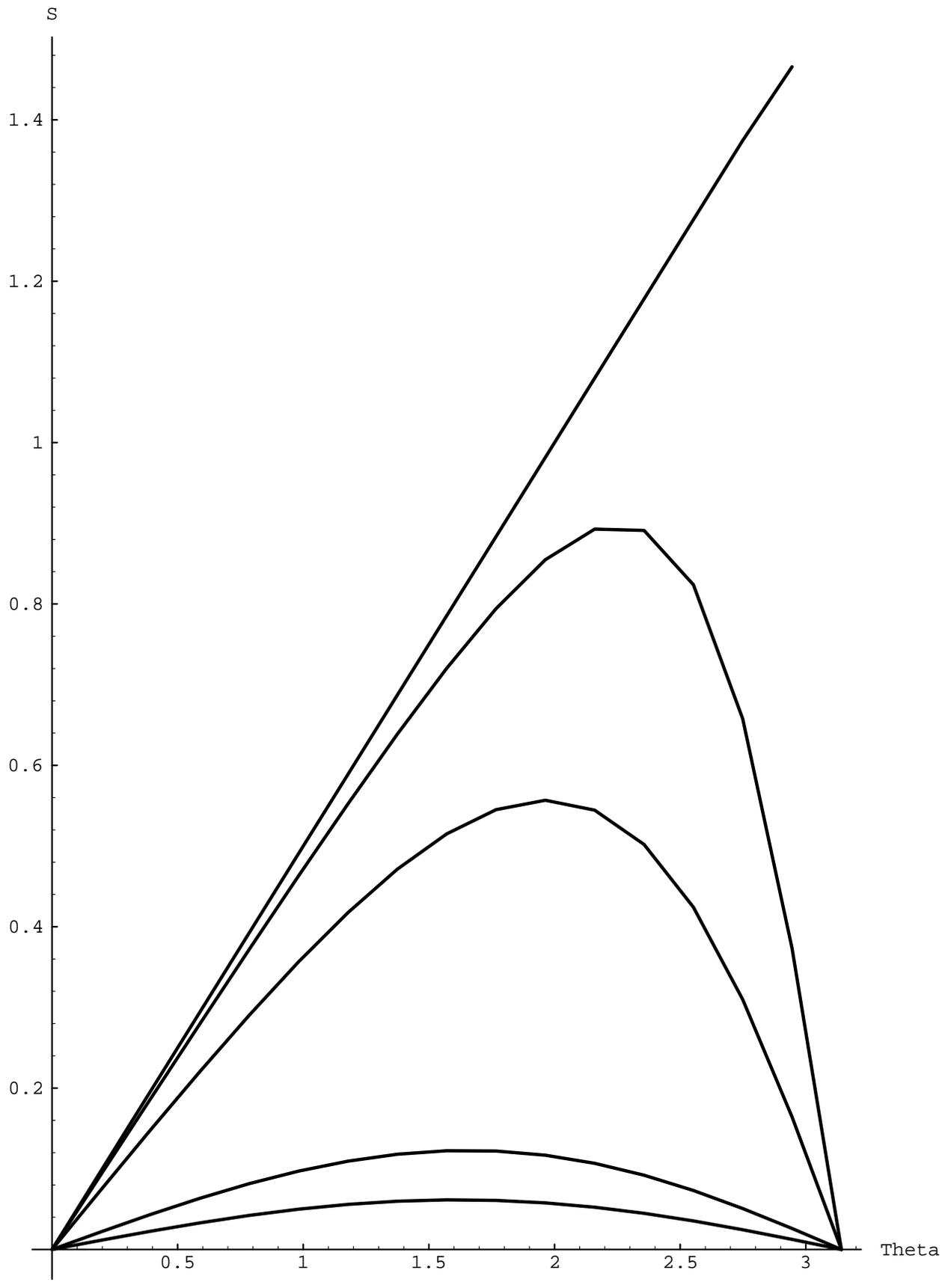
\end{document}